\documentclass[preprint,aps]{revtex4}
\usepackage{graphicx}
\newcommand{\be}{\begin{equation}}
\newcommand{\ee}{\end{equation}}
\newcommand{\ba}{\begin{eqnarray}}
\newcommand{\ea}{\end{eqnarray}}
\newcommand{\p}{\partial}
\newcommand{\f}{\frac}

\begin{document}
\title
{The Effects of Light-Wave Nonstaticity on Accompanying Geometric-Phase Evolutions
 }
\author{Jeong Ryeol Choi\footnote{E-mail: choiardor@hanmail.net } \vspace{0.3cm}}

\affiliation{Department of Nanoengineering, 
Kyonggi University, 
Yeongtong-gu, Suwon,
Gyeonggi-do 16227, Republic of Korea \vspace{0.7cm}}

\begin{abstract}
\indent
Quantum mechanics allows the emergence of nonstatic quantum light waves in the Fock state
even in a transparent medium of which
electromagnetic parameters do not vary over time.
Such wave packets
become broad and narrow
in turn periodically in time
in the quadrature space.
We investigate the effects of
wave nonstaticity arisen in a static environment on the behavior of accompanying geometric
phases in the Fock states.
In this case, the geometric phases appear only when the measure of nonstaticity is not zero and
their time behavior
is deeply related to the measure of nonstaticity.
While the dynamical phases undergo linear decrease over time, the geometric phases
exhibit somewhat oscillatory behavior where the center of oscillation linearly increases.
In particular, if the measure of nonstaticity is sufficiently high, the geometric phases abruptly change
whenever the waves become narrow in the quadrature space.
The understanding for the phase evolution of
nonstatic light
waves is necessary in their technological applications regarding wave modulations.
\\

\indent \vspace{0.0cm} {\bf Keywords}: wave nonstaticity; geometric phase; light wave; Hannay angle;
wave function
\end{abstract}

\maketitle

{\ \ \ } \\
{\bf 1. INTRODUCTION \vspace{0.2cm}} \\
The Berry's seminal discovery \cite{ber} to the appearance of an additional phase evolution in eigenstates of
the Hamiltonian during a slow variation of a quantum system in the parameter space had triggered extensive
research for such extra phases
in both theoretical and experimental spheres.
The Berry phase is a geometrical character of quantum waves, which corresponds to a holonomy
transformation in state space.
It is impossible to gauge out the geometrical character in the phase because
the Berry phase or, in general, the geometric phase is
a gauge invariant.
For this reason, the geometric phase
is non-negligible and, especially, inevitable for analyzing the
transmission of light waves in media
with time-varying parameters since it reflects the geometry of a quantum wave evolution.
It has been proved that the original concept of the Berry's geometric phase can be extended to
more general cases which are nonadiabatic, non-cyclic and/or
non-unitary evolutions of light waves \cite{aha,nmu}.

The geometric phase is a promising research
subject that has been widely investigated with the purpose of
manipulating quantum phases of light waves and controlling their behaviors.
The scientific fields that the geometric phase can be applicable are plentiful: they include
a holonomic quantum computation with geometric gates \cite{cvg},
a stellar interferometry \cite{sti},
the testing of
CPT (charge conjugation, parity, and time reversal) invariance in particle physics \cite{cpt},
entanglement of atoms \cite{gpr1}
analysis of Aharonov-Bohm effect \cite{gpr2}, etc.
Among them, a geometric quantum computation enables us to carry out quantum logic operations by means of
multi-qubit gates, which is
a main
technique for realizing quantum computers \cite{cvg}.

It is assumed that the geometric phase appears for nonstatic
states of a quantum system, whereas
it always vanishes for stationary states \cite{conn}.
As is well known, the ordinary waves
in the Fock state
are static in time, resulting in
no emergence of the geometric phase.
However, if we prepare a quantum wave with time-varying eigenfunctions, there will appear
geometric phases even for a simple situation where
the parameters of the medium do not vary over time.
For instance, the eigenfunctions in  coherent and squeezed states are expressed in terms of time
regardless that the parameters of the media depend on time or not \cite{igp}.
This leads to the appearance of the geometric phase in such states \cite{snb,igp}.

Meanwhile, it was reported from our recent work \cite{nwh} that nonstatic quantum light waves
can also appear in the Fock state in a static environment associated with a transparent medium.
During the time evolution of such wave packets, the waves exhibit a peculiar
behavior as a manifestation of their nonstaticity, which is that they become narrow and
broad in turn periodically in quadrature space.
Subsequently, we also analyzed the mechanism underlain in such a phenomenon from a fundamental
point of view \cite{gnb}.

Even if the environment is static in that case, the periodical time variation of the waves is accompanied by
the evolution of the geometric phases.
This is due to the fact that the
eigenfunctions also vary in time along the nonstaticity of the wave.
We will investigate the characteristics of the geometric phases in the Fock state
arisen in such a situation in this work.
It will be focused on analyzing how the geometric phases evolve in time in relation with
the wave nonstaticity.
We will compare the behavior of the geometric phases with that of the dynamical phases.
The Hannay angle \cite{hha} of the system, which is the classical analogues of the geometric phases, will also be
investigated by utilizing the
relation between them and its physical meanings will be addressed.
Based on the Hannay angle, we can obtain an
insight on the classical geometric structure
of light and its connection with the quantum geometric-phase structure \cite{ilb,hs}.
\\
\\
{\bf 2. DESCRIPTION OF NONSTATIC WAVES \vspace{0.2cm}} \\
To establish the geometric phases for a nonstatic wave,
we first show how to describe nonstatic quantum light waves in a static environment.
The Hamiltonian for a light is given by
\be
\hat{H}= {\hat{p}^2}/{(2\epsilon)} + \epsilon\omega^2 \hat{q}^2 /2, \label{1}
\ee
where $\hat{q}$ is the quadrature operator, $\hat{p} = -i\hbar \p/\p q$,
and $\epsilon$ is the electric permittivity of the medium.
The geometric phases are dependent on the preparation of the wave functions \cite{XX}.
If we consider elementary static wave functions in the Fock state, of which eigenfunctions are
given by
$\langle q|\phi_n \rangle =
({\alpha }/{\pi })^{1/4}({\sqrt{2^{n} n!}})^{-1} H_{n}
\left(\sqrt{\alpha } {q} \right)\exp \left[-{\alpha} {q}^2/{2} \right]$
where $\alpha = \epsilon \omega /{\hbar}$ and $H_n$ are
Hermite polynomials,
the geometric phases do not take place \cite{igp,amo}.
However, for the case of the wave functions whose eigenfunctions are time-dependent,
the geometric phases are nonzero
because the geometric phases are given in terms of the time derivative of the eigenfunctions.
Notice that, for a time-{\it in}dependent Hamiltonian,
including the case regarded here,
there are Schr\"{o}dinger solutions associated with wave nonstaticity as well as
the ones that correspond to static waves.

Instead of $\langle q|\phi_n \rangle$, nonstatic waves in this context can be described
with generalized eigenfunctions of the form \cite{nwh}
\be
\langle q |\Phi_n \rangle =
\left({\f{\beta(t)}{\pi}}\right)^{1/4} \f{1}{\sqrt{2^n
n!}} H_n \left( \sqrt{\beta(t)} q \right) \exp \left[
- \f{\beta(t)}{2} \left(1-i\f{\dot{f}(t)}{2\omega}\right)q^2
\right], \label{3}
\ee
where $\beta(t) = {\epsilon\omega}/{[\hbar f(t)]}$
and $f(t)$ is a time function which is given by
\be
f(t) = A \sin^2 \tilde{\varphi}(t)+ B \cos^2 \tilde{\varphi}(t) +
C \sin [2\tilde{\varphi}(t)], \label{12} \ee
under an auxiliary condition, 
$
AB-C^2 = 1 \label{13}
$
with $AB \geq 1$,
while $\tilde{\varphi}(t)=\omega (t-t_0) +\varphi$ whereas
$\varphi$ is a real constant.
It is now possible to establish time-dependent wave functions
associated with the nonstatic wave in terms of $\langle q |\Phi_n \rangle$, such that \cite{nwh}
\be
\langle q |\Psi_n(t) \rangle = \langle q |\Phi_n(t) \rangle
 \exp \bigg[{-i\omega (n+1/2) \int_{t_0}^t f^{-1} (t') dt'} + i\gamma_n (t_0) \bigg], \label{2}
\ee
where $\gamma_n (t_0)$ are phases at $t_0$.
We note that the wave functions
given above satisfy
the Schr\"{o}dinger equation associated with the Hamiltonian represented
in Eq. (\ref{1}),
and $f(t)$ used here follows the nonlinear differential equation of the form
$
\ddot{f} - {(\dot{f})^2}/({2f}) + 2\omega^2
\left(f- {1}/{f}\right) =0. \label{4}
$

\begin{figure}
\centering
\includegraphics[keepaspectratio=true]{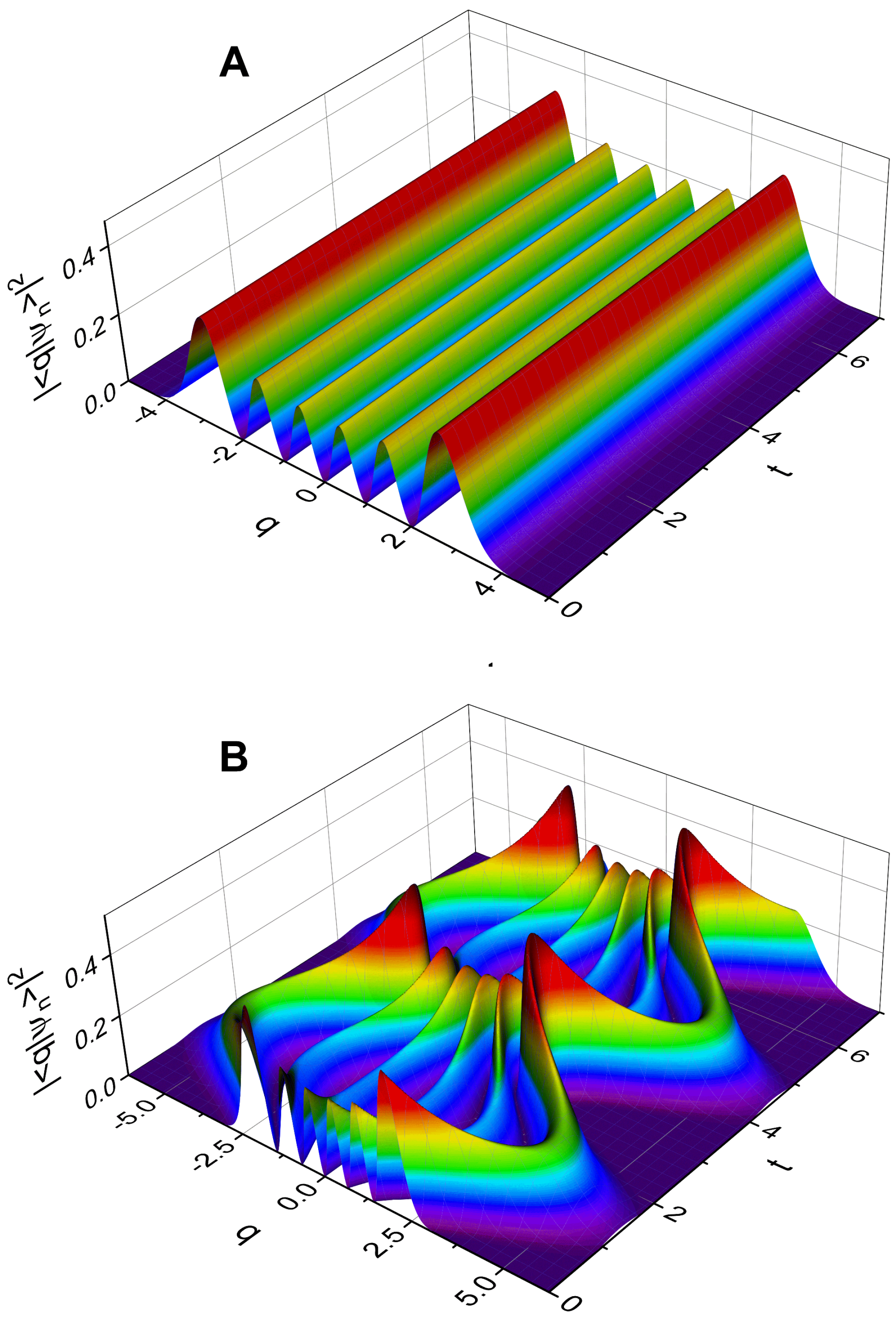}
\caption{\label{Fig1} Comparison between probability densities for static ({\bf A}) and nonstatic ({\bf B}) wave packets.
We have chosen the time function $f(t)$ for {\bf A} as 1 and for {\bf B}
as Eq. (\ref{12}) with $A=2.5$ and $B=0.5$.
We take only positive values for $C$ throughout all figures in this work for convenience.
Then, the value of $C$ is automatically determined from
$A$ and $B$ through the auxiliary condition given below Eq. (\ref{12}).
Other values that we have chosen are $n=5$, $\omega=1$, $\epsilon=1$, $\hbar=1$, $t_0=0$, and $\varphi=0$.
All variables are chosen to be dimensionless for convenience; this rule will also be applied to
subsequent figures.
For the nonstatic case, the probability density undergos cyclic evolution with time period
$T=\pi/\omega$, i.e., the width of the wave packet becomes broad and narrow in turn over time.}
\end{figure}

We will restrict the classical angle within the range $-\pi/2 \leq \varphi < \pi/2$,
because the research in this range is enough owing to the fact that Eq. (\ref{12}) is a periodic function
with the angle period $\pi$.
For $A=B=1$ and $C=0$, Eq. (\ref{12}) reduces to $f(t) =1$ which corresponds to the case
that gives
static wave functions; All other choices for the set of $A$ and $B$ give nonstatic wave functions.
We have compared static and nonstatic wave packets in Fig. 1.
The nonstatic wave packets shown in Fig. 1(B) vary periodically over time.
The degree of such time variation caused by
nonstaticity is determined by the quantitative measure of nonstaticity.
For detailed expression of the nonstaticity measure, see Appendix A.
The nonstatic-wave packets described up until now  will
be used in order to investigate the geometric phases in the subsequent
subsection.
\\
\\
{\bf 3. GEOMETRIC PHASE \vspace{0.2cm}} \\
The geometric phases are examples of holonomy which gives additional phase evolutions
of the quantum wave
over time.
The development of the geometric phases for one-dimensional
simple wave description
offers essential ideas which enable us to demonstrate
topological features in quantum mechanics.
The geometric phases for the nonstatic waves
can be evaluated from \cite{nc3,nc4}
\be
\gamma_{G,n}(t) = \int_{t_0}^t \langle\Phi_n(t')
|i\frac{\partial}{\partial t'}| \Phi_n(t') \rangle dt' +\gamma_{G,n}(t_0).
\label{5} \\
\ee
These are parts of the phases of quantum wave functions at time $t$, which have geometric origin.
On one hand, there is a concept of the geometric phase whose definition is a little different:
it is the geometrical part of the phase acquired during only one cycle
evolution of the eigenstate through a closed path in the circuit
\cite{snb}.
In what follow, we will use the former concept of the geometric phase associated with Eq. (\ref{5}) throughout this paper.
We assume that the initial phases are zero for convenience from now on: $\gamma_{G,n}(t_0)=0$.
It is possible to
evaluate Eq. (\ref{5}) by using Eq.
(\ref{3}) with the consideration of
$f(t)$ given in Eq. (\ref{12}). Hence, we have
(see METHODS section which is the last section)
\be
\gamma_{G,n}(t) = \f{1}{2} \left( n+\f{1}{2} \right)
\{(A+B)\omega (t-t_0) -2 [\tan^{-1} Z(t) - \tan^{-1} Z(t_0)+G(t) ]\}, \label{14-1}
\ee
for $t\geq t_0$, where $Z(\tau)= C+A\tan[\omega (\tau-t_0) +\varphi]$,
and $G(t)$ is a time function that is expressed in terms of the unit step
function (Heaviside step function) $u[t]$ as
$G(t)=\pi \sum_{m=0}^{\infty}u[t-t_0-(2m+1)\pi/(2\omega)+\varphi/\omega]$.
If we regard the periodical discontinuities of tangent functions in Eq. (\ref{14-1}) with a period of $\pi$,
$G(t)$ is necessary in order to compensate them in a way that
$\gamma_{G,n}(t)$ become continuous functions.
We note that Eq. (\ref{14-1}) holds within the
considered range for $\varphi$, $-\pi/2 \leq \varphi < \pi/2$ (see the previous subsection for this range).
Thus we have obtained the geometric phases for the nonstatic light waves.
The formula of the geometric phases given above may provide a deeper insight for the understanding of the
nature of the nonstatic waves.

The dynamical phases can also be derived from their definition using the same wave functions
and are given by
$
\gamma_{D,n}(t) = -(1/2)( n+{1}/{2} )(A+B)\omega (t-t_0)
$
(see METHODS section).
If we regard that the measure of nonstaticity shown
in Appendix A
is nearly proportional to
$A+B$ provided that
$A+B \gg 4$,
$\gamma_{D,n}(t)$ at a certain time is linearly proportional to
the measure of nonstaticity for highly nonstatic waves.

\begin{figure}
\centering
\includegraphics[keepaspectratio=true]{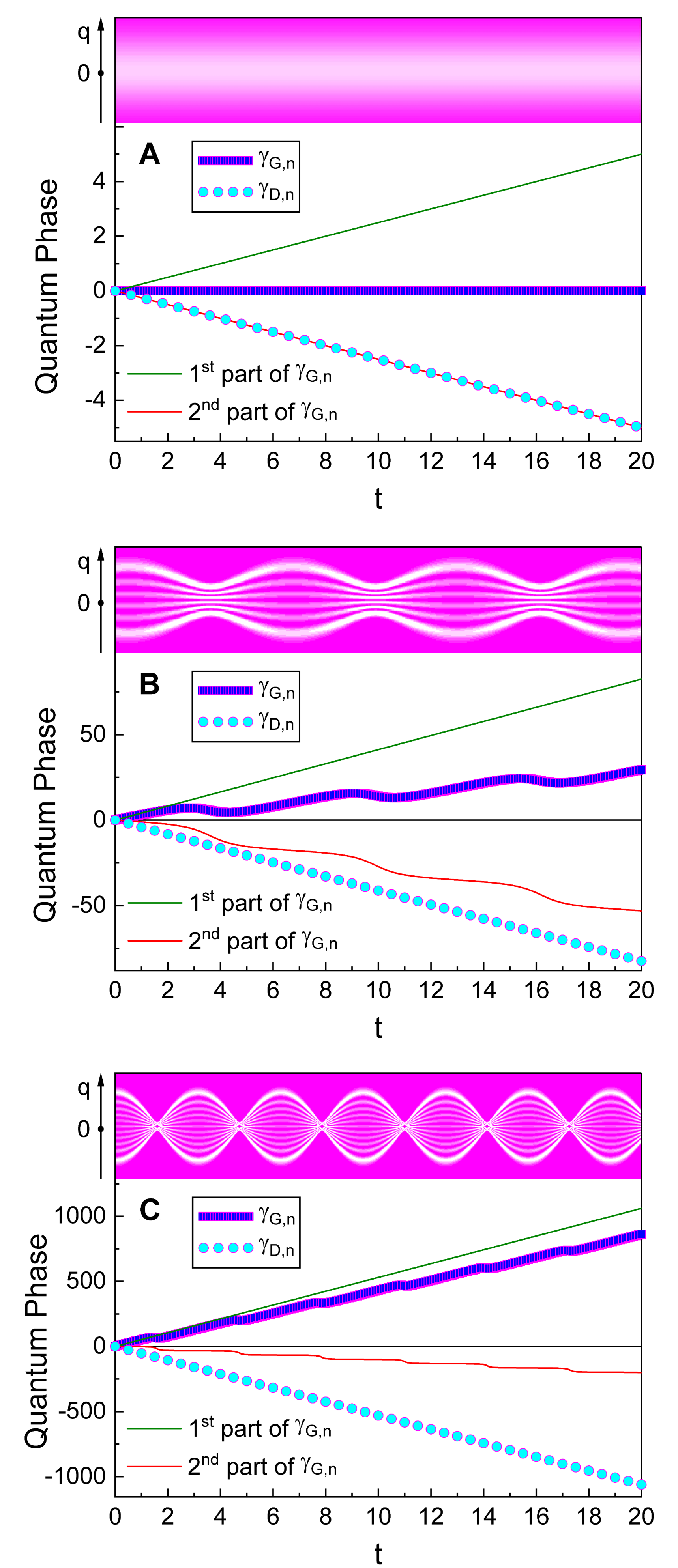}
\caption{\label{Fig2}
Time evolution of the geometric phase in addition to the dynamical phase
for several different values of the parameters.
The values of ($A$, $B$, $n$, $\omega$) are ($1$, $1$, $0$, $0.5$) for {\bf A},
($0.5$, $2.5$, $5$, $0.5$) for {\bf B}, and ($0.1$, $10.0$, $10$, $1$) for {\bf C}.
We have chosen other parameters as $t_0=0$, and $\varphi=0$.
The pink-white graphics in the upper part of the panels are the
time evolution of the corresponding probability density.
The measure of nonstaticity is 0.00 for {\bf A}, 0.79 for {\bf B}, and 3.50 for {\bf C}.}
\end{figure}

Whereas the dynamical phases linearly decrease over time from their initial values,
the time behavior of the geometric phases is not so simple.
Let us divide the geometric phase in Eq. (\ref{14-1}) into two parts for the convenience of analyses.
We call the term that involves $(A+B)$ as the first part
and the remaining terms the second part.
We readily confirm that the first part, which increases in a monotonic manner over time,
exactly cancels the dynamical phase.
Hence, the second part of the geometric phase is the same as the total phase of the wave.
However, it may be not so easy to completely estimate the
evolution of the geometric phases since
the second part
is somewhat intricate.

To understand the overall time behavior
of the geometric phases,
we plotted the evolution of the geometric phase in Fig. 2 together with the dynamical phase
for several different values of parameters.
Let us first examine the effects of $A$ and $B$ on the geometric phase.
For a trivial case where $A=B=1$ and $C=0$, which corresponds to Fig. 2(A), $f(t)$ becomes
unity and the eigenfunctions, Eq. (\ref{3}), reduce to time-{\it in}dependent ones
as have seen from the previous subsection.
As a consequence, the geometric phases result in
$
\gamma_{G,n}(t) = 0 \label{10}
$
whereas the dynamical phases become
$
\gamma_{D,n}(t) = -( n+{1}/{2} )\omega (t-t_0)  . \label{11}
$
Hence, the geometric phases do not appear in this case as it should be,
while the dynamical parts
are the well known formula.
We confirm from Fig. 2(A), which is the case of the static wave,
that
the second part of the geometric phases linearly decreases over time, leading exact cancelling with
the first part.
In addition, the second part of the geometric phase is the same as the dynamical phase only when $A=B=1$.

\begin{figure}
\centering
\includegraphics[keepaspectratio=true]{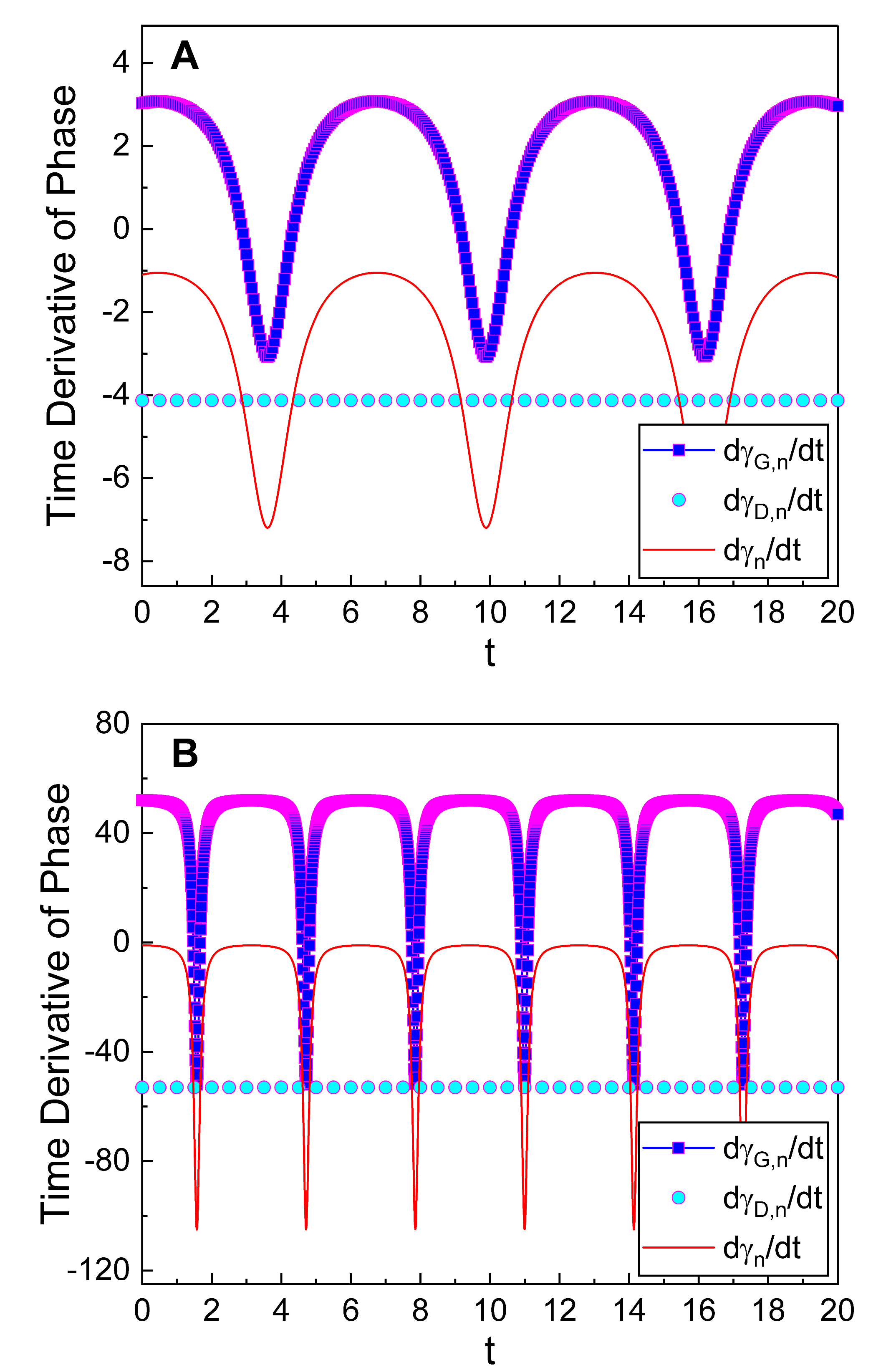}
\caption{\label{Fig3} Behavior for the time derivative of the
geometric phase $d\gamma_{G,n}(t)/dt$, the dynamical phase $d\gamma_{D,n}(t)/dt$, and the
 total phase $d\gamma_{n}(t)/dt$.
All chosen parameters for {\bf A} and {\bf B} are the same as
those for {\bf B} and {\bf C} in Fig. 2,
 respectively.
}
\end{figure}

If at least one of $A$ and $B$ is not unity, it becomes the case of a nonstatic wave
as shown in Figs. 2(B) and 2(C).
Then, the width of the corresponding probability density
periodically varies over time with the period of $T=\pi/\omega$.
Hence, the frequency in its periodic change is large when $\omega$ is high.
We can confirm from Figs. 2(B) and 2(C) that the second part of the geometric phase changes depending
on the width of the wave packet.
In a moment when the width is large, the second part nearly monotonically decreases over time.
However, when the width is small, the second part somewhat abruptly decrease.
Thus, the geometric phase, which is the addition of the first and the second parts in the figure,
varies according to the evolution of the wave packet.
We can more concretely demonstrate these situations from Fig. 3 which shows that the
time derivative of the geometric phase abruptly drops when the width of the wave packet is
narrow. By the way, the dynamical phases vary in a monotonic manner in all situations.
In this way, the system will pick up a memory of its time evolution in the form of the geometric phase which will
contribute to an observable shift of the phase in the wave function.

\begin{figure}
\centering
\includegraphics[keepaspectratio=true]{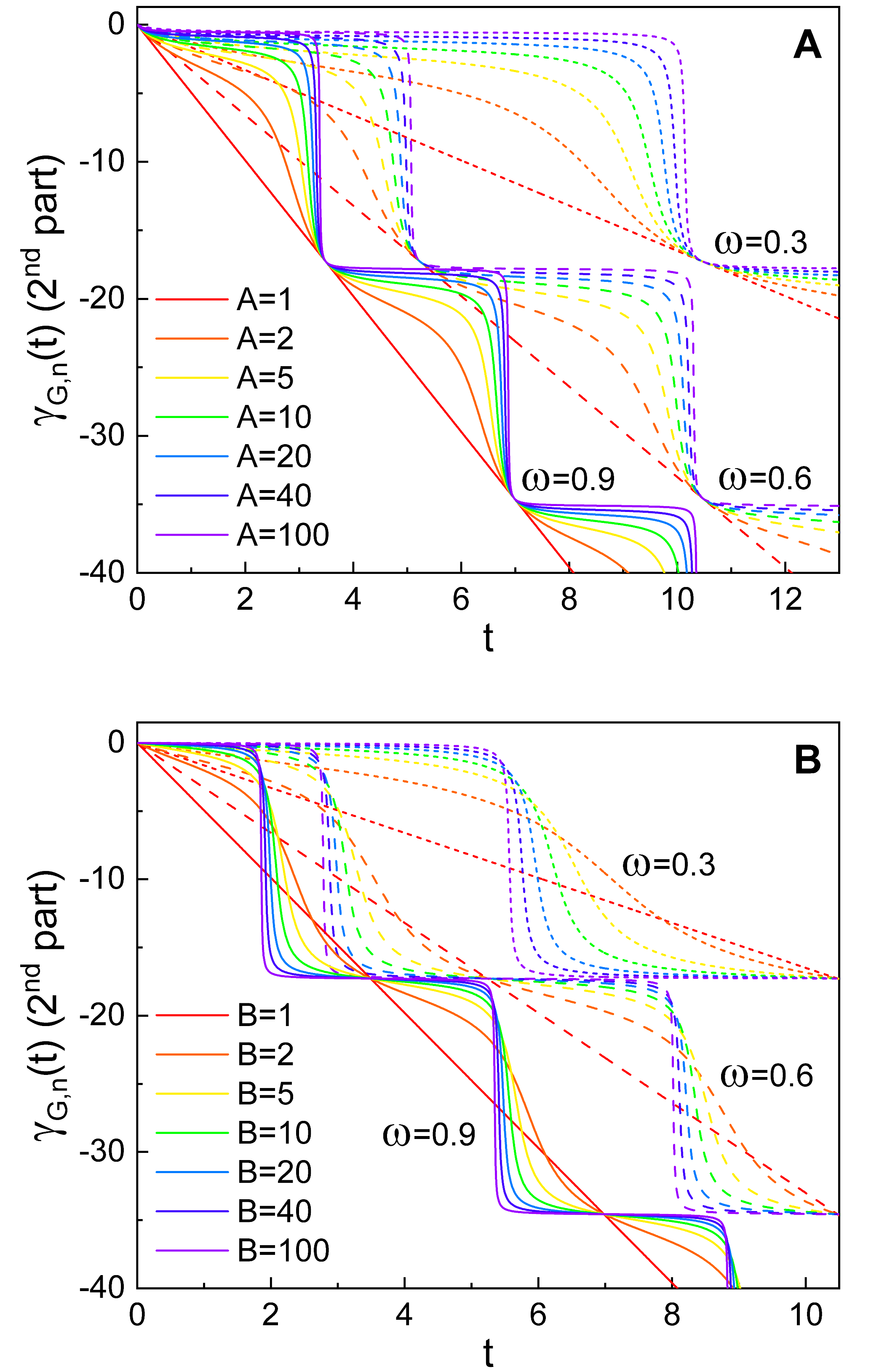}
\caption{\label{Fig4}
The evolution of the second part of the geometric phase
for several values of $A$ ({\bf A}) and $B$ ({\bf B}),
where $B=1$ for {\bf A}, $A=1$ for {\bf B},
$n=5$, $\varphi=0$, and $t_0=0$.
The conventions for colors designated for solid lines in the legends are also applied to the dashed
and the short dashed lines within the figure panels.
The measure of nonstaticities in turn from red to violet curve are 0.00, 0.79, 2.00, 3.82, 7.39, 14.48,
and 35.70 for both {\bf A} and {\bf B}.
}
\end{figure}

The evolution of the second part of the geometric phase is illustrated in Fig. 4.
Although
the first part of the geometric phase becomes large as the value of $A+B$
(or the measure of nonstaticity)
increases, the envelope of the second part is not so significantly affected by the values of
$A$ and $B$.
Actually, the gradient of the envelope of the second part
is irrelevant
to the measure of nonstaticity; such a gradient
is determined by $\omega$ instead.
The envelope of the second part
decreases more rapidly as $\omega$ grows, whereas the
first part increases more rapidly at the same condition.

Figure 4 shows that, when both $A$ and $B$ are unity, the second parts
linearly decrease as time goes by.
On the other hand, if $A$ and/or $B$ deviate from the unity, the gradient
in the phase evolution is not constant over time.
We can confirm from Fig 4(A) that the second parts
abruptly drop when $\omega t$ is $\pi$, $2\pi$, $3\pi$, etc.
provided that $A$ is very large while $B$ is unity; however, except for these moments, the second parts
almost do not vary over time.
As a consequence, the bottom envelope of the second parts
is identical to the standard
value which is $-( n+{1}/{2} )\omega (t-t_0)$.
Similar behaviors in the phase evolution can also be seen from specific curves in Fig. 4(B),
which correspond to the case where $B$ is very large while $A$ is unity;
in this case, the second parts drop when $\omega t$ is $\pi/2$, $3\pi/2$, $5\pi/2$, etc.

The dependence of the geometric-phase evolution on $\varphi$ within the considered region
$-0.5\pi \leq \varphi < 0.5\pi$ is shown in Fig. 5.
All geometric phases in the figure start from zero, but the interval of phase
between the adjacent geometric phases is $\Delta \varphi = 0.15 \pi$.
Because the geometric phase periodically varies over $\varphi$ where the period of such a variation is $\pi$,
it is not difficult to know the pattern of the geometric phase outside the considered region for $\varphi$.
More precisely speaking, the value of the Berry phase
for $\varphi=\pi$ is exactly the same as that for $\varphi=0$; the value of the Berry phase
for $\varphi=1.15\pi$ is exactly the same as that for $\varphi=0.15\pi$, etc.

\begin{figure}
\centering
\includegraphics[keepaspectratio=true]{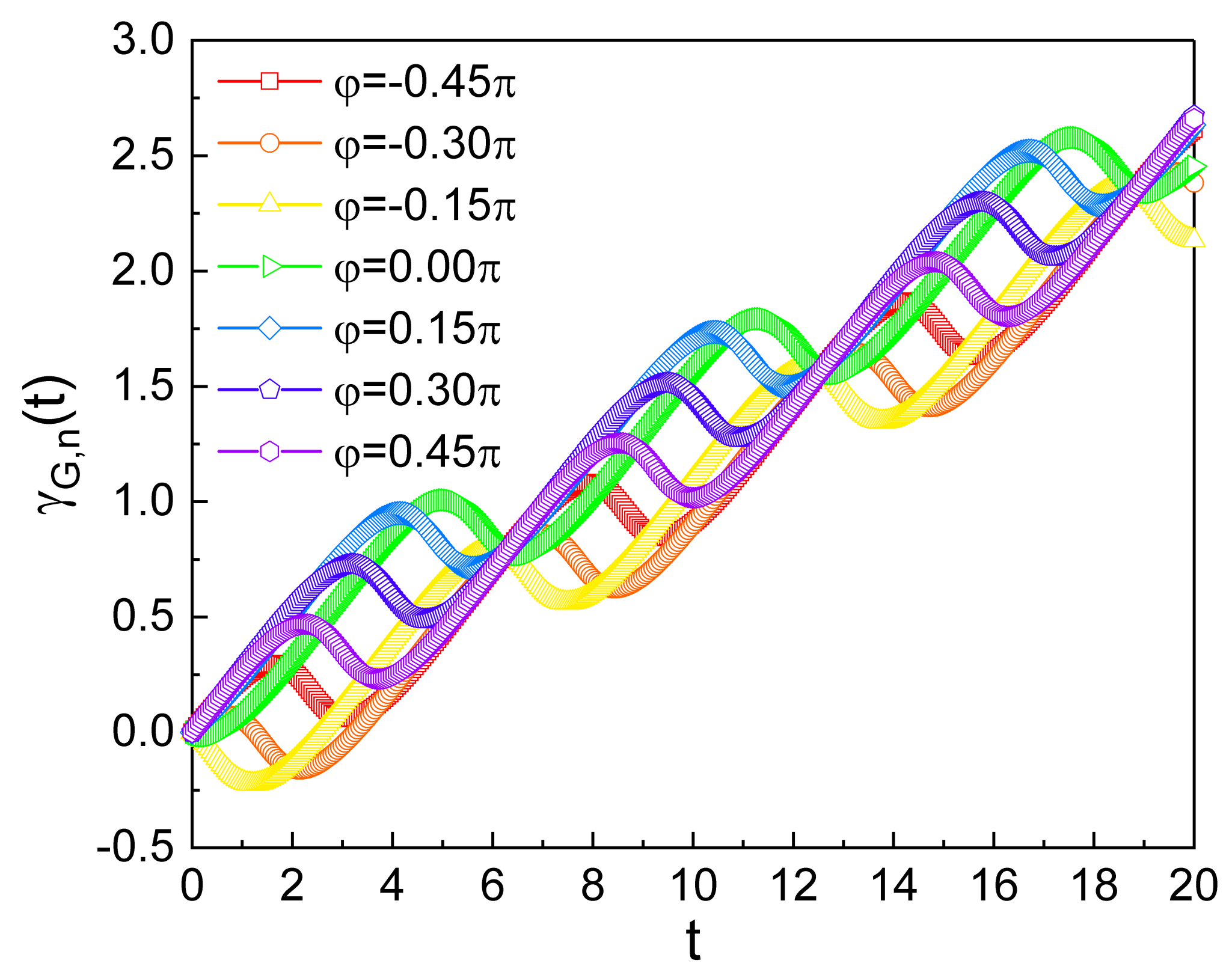}
\caption{\label{Fig4} Time evolution of the geometric phase for various values of $\varphi$.
We have chosen parameters as $A=2.5$, $B=0.5$, $n=0$, $\omega=0.5$, and $t_0=0$.
}
\end{figure}

The overall phases of the waves are given as
$
\gamma_n(t) = \gamma_{G,n}(t) + \gamma_{D,n}(t) .  \label{7}
$
Hence, by adding our two results for geometric and dynamical phases,
we have the total phases such that
$
\gamma_n(t) = -\omega ( n+ {1}/{2} )\int_{t_0}^t f^{-1}(t') dt' . \label{6-1}
$
These formulae are exactly the same as the phases having appeared in the wave functions, Eq. (\ref{2}),
under the condition that the initial phases are zero, $\gamma_n(t_0)=0$.

Hannay confirmed that the phase similar to the geometric phase also appears in the classical
domain as an analogy with the geometric phase \cite{hha}.
This is a geometrical angle for a classical wave which allows to estimate its holonomy effect
from the corresponding geometrical interpretation.
This is an important theoretical concept and, hence, it may be instructive
to see the Hannay angle of the system.
For an integrable classical system, action-angle variables theory at the semiclassical level
from torus quantization gives the fact that
the Hannay angle $\Theta_H (t)$ is related to the geometric phase as
$
\Theta_H (t)= - {\partial \gamma_{G,n}(t)}/{\partial n} \label{27}
$ \cite{bjp,xbw}.
From this, we can easily confirm that the Hannay angle in this case is of the form
$
\Theta_H(t) = - 2\gamma_{G,0}(t). \label{28}
$
This natural relation reveals that the evolution of the classical additional angle
can be represented in terms of
the geometric phase associated with the zero-point quantum wave packet.
This simple clear picture provides a significant geometric meaning for unification of the gauge
structure of the wave propagation for quantum and classical mechanics,
leading in effect to Bohr's correspondence principle \cite{ggb}.
If we develop a method to generate
the generalized wave packet given by Eq. (\ref{2}) in the future,
the demonstration of this consequence may be possible by measuring Hannay angle using an
averaging technique introduced in Refs. \cite{gse,gse2,gse3}.
The reason why the Hannay angle takes place in the classical system is
that the structure of the quantum Hilbert space
is quite the same as the classical phase space \cite{dch,yia}.
\\
\\
\\
\\
{\bf 4. CONCLUSION
\vspace{0.2cm}} \\
Ordinary quantum light waves, which propagate through a transparent medium in which electromagnetic
parameters do not vary,
are static in time,
resulting in no appearance of
the geometric phase in Fock states.
However, if one or both of the parameters $A$ and $B$ deviates from unity, the waves
in the Fock states in that medium become
nonstatic and, as a consequence, the geometric phases emerge.
Not only the appearance of such nonstatic waves but also the characteristics of the resultant
geometric phases may be
noteworthy \cite{nwh}.
We have analyzed the influence of the wave nonstaticity
on the evolution of the geometric phases in such a case.

The geometric phases of the light waves exhibit periodical oscillatory
behavior with the period of $\omega T = \pi$, where the center of such
oscillation linearly increases over time.
On the other hand, the dynamical phases always show linear decrease.
Because the scale of the geometric phases is smaller than that of the dynamical phases,
the total phases evolve toward opposite
direction over time.
If the measure of nonstaticity is high,
both the geometric and the dynamical phases rapidly evolve in time.
Although the geometric phases increase on the whole,
they periodically drop with the angle period of $\pi$ provided that the
measure of nonstaticity is sufficiently high.
Such a variation of the geometric phases
is quite
significant when the measure of nonstaticity
is extremely large.

We have shown that Hannay angle, which is the classical analogue of the geometric phase, is represented
in terms of the geometric phase
associated to the zero-point wave function.
This elegant outcome
shows a unified picture of the interpretation of the geometric character of light waves
in the quantum and the classical regime.
This
not only bridges the quantum and classical world,
but can also be extended to more generalized quantum light waves, such as the light in
a squeezed state and the Gaussian wave packet propagating in time-varying media
\cite{jli,jli2}.
The understanding of the evolution of the geometric phases for
nonstatic quantum light waves
in a static environment is necessary in
quantum optics,
especially in relation with wave modulations.
Practical utilization of the geometric phases in diverse scientific areas may be available under
the fundamental
knowledge associated with the behavior of quantal phases.

On one hand, the geometric phase of quantum systems which evolve
in a non-unitary way \cite{nuw} and its relation with wave
nonstaticity as well as accompanying non-unitary effects such as
decoherence and dissipation may be worthwhile to be explored in a next research.
Kinematic approach to the geometric phase of non-unitarily evolving systems
is important in the pursuing of robustness of geometric quantum computation \cite{ka1,ka2,ka3}.
\\
\\
{\bf 5. METHODS \vspace{0.2cm}} \\
Let us see how to evaluate the geometric and the dynamical phases.
We first derive the geometric phases.
From a minor computation in the configuration space after
inserting Eq. (\ref{3}) into Eq. (\ref{5}), we have under the condition $\gamma_{G,n}(t_0)=0$:
\be
\gamma_{G,n}(t) = \f{1}{2} \left( n+\f{1}{2} \right)
\Gamma_{G}, \label{14} \\
\ee
where
\be
\Gamma_{G} = \omega[g_1(t)-g_2(t)]+ \f{g_3(t)}{4\omega}, \label{16}   \\
\ee
with
\ba
g_1(t) &=& \int_{t_0}^t f(t') dt' , \label{29} \\
g_2(t) &=& \int_{t_0}^t \f{1}{f(t')} dt' , \label{30} \\
g_3(t) &=& \int_{t_0}^t \f{[\dot{f}(t')]^2}{f(t')} dt' . \label{31}
\ea

Straightforward evaluations of $g_i(t)~(i=1,2,3)$ using Eq. (\ref{12}) yield
\be
g_i(t) = G_i(t) - G_i(t_0), \label{18} \\
\ee
for $t_0 \leq t<t_0+\pi/(2\omega)-\varphi/\omega$, where
\ba
G_1(\tau) &=& \f{1}{4\omega} \{ 2 (A+B)\omega \tau -(A-B)\sin\{2[\omega(\tau-t_0)+\varphi]\} \nonumber \\
& &-2C \cos\{2[\omega(\tau-t_0)+\varphi]\} \}, \label{19}  \\
G_2(\tau) &=& \f{1}{\omega}\tan^{-1} \{ C+A\tan[\omega (\tau-t_0) +\varphi] \}, \label{20}  \\
G_3(\tau) &=& \omega \{ 2(A+B)\omega \tau
 +(A-B)\sin\{ 2[\omega (\tau-t_0)+\varphi] \} \nonumber \\
 & &+2C\cos\{ 2[\omega (\tau-t_0)+\varphi] \} - 4\tan^{-1}\{ C+A\tan[\omega (\tau-t_0) +\varphi] \} \}.
  \label{21}
\ea
By readjusting Eq. (\ref{16}) with Eqs. (\ref{18})-(\ref{21}), we have
\be
\Gamma_{G} = F_G(t)-F_G(t_0),  \label{22}
\ee
where $F_G(\tau)$ is given by
\be
F_G(\tau) = (A+B)\omega \tau -2 \tan^{-1} \{ C+A\tan[\omega (\tau-t_0) +\varphi] \}.
 \label{24-1}
\ee
Equation (\ref{22}) hold for the region $t_0 \leq t<t_0+\pi/(2\omega)-\varphi/\omega$,
because we have considered $t \geq t_0$ and there is the first discontinuity in the tangent function at $\pi/2$.
If we want to extend the expression in this equation to the whole region ($t \geq t_0$) that we have considered,
it is necessary to compensate Eq. (\ref{22}) by the unit step function $u[t]$ to be
\be
\Gamma_{G} = F_G(t)-F_G(t_0)-2\pi \sum_{m=0}^{\infty}u[t-t_0-(2m+1)\pi/(2\omega)+\varphi/\omega].  \label{b22}
\ee
By rearranging Eq. (\ref{14}) with Eqs. (\ref{b22}) and (\ref{24-1}), we can easily have the
formula of geometric phases which are given in Eq. (\ref{14-1}) in the text.

Now we evaluate the dynamical phases. The definition of the dynamical phases are given by
\be
\gamma_{D,n}(t) = - \f{1}{\hbar} \int_{t_0}^t \langle\Phi_n(t') |
\hat{H}(\hat{q},\hat{p},t')| \Phi_n(t')\rangle dt' + \gamma_{D,n}(t_0). \label{6}
\ee
We also assume that $\gamma_{D,n}(t_0)=0$ like the case of the geometrical part.
Using the Hamiltonian of the simple harmonic oscillator and the Fock states given in Eq. (\ref{3}),
we can evaluate Eq. (\ref{6}) and this results in
\be
\gamma_{D,n}(t) = \f{1}{2} \left( n+\f{1}{2} \right)
\Gamma_{D} , \label{15}
\ee
where
\be
\Gamma_{D} = -\omega[g_1(t)+g_2(t)]- \f{g_3(t)}{4\omega}. \label{17}
\ee
Using Eqs. (\ref{18})-(\ref{21}), we have
\be
\Gamma_{D} =  F_D(t)-F_D(t_0), \label{23}
\ee
where
\be
F_D(\tau) = -(A+B)\omega \tau . \label{25}
\ee
A minor readjustment of the above equations gives
\be
\gamma_{D,n}(t) = \f{1}{2} \left( n+\f{1}{2} \right)
[F_D(t)-F_D(t_0)] . \label{15-1}
\ee
We can easily show that this is the same formula of the dynamical phases given in the text.
\appendix
\section{Measure of nonstaticity}
In our previous paper \cite{nwh}, we have defined the quantitative measure of nonstaticity.
In the Fock state, it is given by \cite{nwh}
\be
D_{\rm F} = \f{\sqrt{(A+B)^2-4}}{2\sqrt{2}}.  \label{}
\ee
Usually, the nonstatic properties of the light wave become significant as this measure increases.
\\

\end{document}